\documentclass[aps,prb,twocolumn,superscriptaddress]{revtex4-2}
\usepackage{graphics,graphicx, array, afterpage}
\usepackage{qcircuit,amsmath}
\usepackage{grffile}
\usepackage[export]{adjustbox}
\usepackage{lineno}
\usepackage{xcolor}
\usepackage{longtable}
\usepackage{braket}
\usepackage{array}
\usepackage{multirow}
\usepackage{enumerate}
\usepackage{amsmath}
\usepackage{amssymb}
\usepackage{amsthm}
\usepackage{times}
\usepackage{multirow}

\definecolor{LinkColor}{RGB}{0,0,128} 
\usepackage[colorlinks, citecolor=blue]{hyperref}
\hypersetup{linkcolor=LinkColor,urlcolor=LinkColor,citecolor=LinkColor,pdfstartview={FitH},urlcolor=LinkColor}

\usepackage[utf8]{inputenc}
\usepackage{color}
\usepackage{tabularx}
\usepackage{algorithm}
\usepackage{algpseudocode}
\usepackage{graphbox}
\usepackage{amsfonts}
\usepackage{dcolumn}
\usepackage{bm}
\usepackage{epstopdf}

\begin{document}

\preprint{APS/123-QED}
\title{Fate of pseudo mobility-edge and multiple states in non-Hermitian Wannier-Stark lattice}

\author{Yu-Jun Zhao}
\thanks{These two authors contributed equally to this work.}
\affiliation{School of Physics and Optoelectronics, Xiangtan University, Xiangtan 411105, China}
\affiliation{Institute for Quantum Science and Technology, Shanghai University, Shanghai 200444, China}

\author{Han-Ze Li}
\thanks{These two authors contributed equally to this work.}
\affiliation{Institute for Quantum Science and Technology, Shanghai University, Shanghai 200444, China}

\author{Xuyang Huang}
\affiliation{School of Physics and Optoelectronics, Xiangtan University, Xiangtan 411105, China}
\affiliation{Institute for Quantum Science and Technology, Shanghai University, Shanghai 200444, China}

\author{Shan-Zhong Li}
\email{szhongli@m.scnu.edu.cn}
\affiliation{Key Laboratory of Atomic and Subatomic Structure and Quantum Control (Ministry of Education),
Guangdong Basic Research Center of Excellence for Structure and Fundamental Interactions of Matter,
School of Physics, South China Normal University, Guangzhou 510006, China}
\affiliation{Guangdong Provincial Key Laboratory of Quantum Engineering and Quantum Materials,
Guangdong-Hong Kong Joint Laboratory of Quantum Matter,
Frontier Research Institute for Physics, South China Normal University, Guangzhou 510006, China}

\author{Jian-Xin Zhong}
\email{jxzhong@shu.edu.cn}
\affiliation{Institute for Quantum Science and Technology, Shanghai University, Shanghai 200444, China}
\affiliation{School of Physics and Optoelectronics, Xiangtan University, Xiangtan 411105, China}
\date{\today}

\begin{abstract}

The interaction between non-reciprocity and disorder-free localization has emerged as a fascinating open question. Here, we explore the effects of pseudo mobility edges (MEs) along with different types of eigenstates in a one-dimensional (1D) lattice subjected to a non-reciprocal finite-height Wannier-Stark ladder. Utilizing the transfer matrix method, we analytically investigate the pseudo mobility edges under non-reciprocity, which accurately describe the boundary between ergodic and non-ergodic states. 
The ergodic states, under non-reciprocity, form topological point gaps in the complex plane, with the corresponding eigenstates localized at the boundaries. 
The localization of mixed states induced by the skin effect and Wannier-Stark ladder is further amplified under non-reciprocity. Through similarity transformations, the fate of multiple eigenstates under non-reciprocal transitions can be captured. Finally, we use wave packet dynamics as a means to detect these emerging states. Our findings broaden the understanding of disorder-free localization in non-Hermitian systems.
\end{abstract}

\maketitle
\section{Introduction} 
Ergodicity principle, the cornerstone to statistical physics, breaks down in disordered quantum systems, leading to localized states where the system fails to explore its entire phase space, as seen in Anderson localization~\cite{1}. Remarkably, a critical energy threshold, known as the single-particle ME~\cite{2,5,6,7,zhong2,zhong1,Gao}, separates these localized states from extended states, with localization typically occurring at lower energies. Initially, MEs identified in disordered systems have been essential in understanding electronic transport and thermalization. While the ME rarely survives in 1D disordered systems due to dimensional constraints, it can emerge in 1D quasiperiodic systems. The most prominent example of a quasiperiodic model is the Aubry-Andr{\'e}-Harper (AAH) model~\cite{3,4}, where the localization transition can be analytically determined by its self-duality symmetry. Furthermore, certain modifications of the AAH model~\cite{9,10,11,12,13}, which disrupt this self-duality, reveal the presence of MEs. These MEs serve as a crucial energy level that distinguish between extended and localized eigenstates, providing deeper insights into the dynamics of quasiperiodic systems and expanding the understanding of localization phenomena beyond traditional disordered models.

Nevertheless, the framework of a ME in 1D disorder-free localization systems has been rarely discussed. Very recently, Ref.~\cite{8} conclusively demonstrated the absence of a ME under disorder-free localization in systems with a Wannier-Stark linear field. In such a system, the Wannier-Stark ladder~\cite{14,15,16,17,18,19, shuo} describes the energy ladder structure that emerges when a particle in a 1D lattice is subjected to a gradient field. This structure leads to an hyper-exponential decay of the wave function. In the thermodynamic limit, the accumulation of the potential gradient across the lattice increases the energy differences between different positions, thereby eliminating the possibility of energy resonance. As a result, all states become localized, and the particle is prevented from freely propagating through the lattice, even when the gradient field is very slight, leading to a scenario where no ME exists. In this context, the Lyapunov exponents (LE) defined by Avila's global theory~\cite{A1,A2,8} fails to effectively distinguish between localized and extended states. Interestingly, while a true disorder-free ME cannot be discussed, in a finite-height Wannier-Stark linear field modified 1D lattice~\cite{wei}, the LE still has a well-defined role. This leads to a pseudo ME that distinguishes between ergodic, weakly ergodic, and non-ergodic states~\cite{wei, jiang}. 

Interestingly, in non-Hermitian systems~\cite{PhysRevB.97.121401,33,34,35,36,37,38,39, longwenzhou, xufeng, zechuan}, some anomalous localization behaviors have emerged, with the most notable example being the non-Hermitian skin effect (NHSE)~\cite{20,21,22,23,24}. This phenomenon describes the localization of a large number of eigenstates at the boundaries under open boundary conditions (OBCs). In addition to the boundary-sensitive NHSE, numerous novel forms of the NHSE have emerged, such as higher-order NHSE~\cite{PhysRevB.102.205118}, geometry-dependent NHSE~\cite{Zhou2023} , and occupation number-dependent NHSE in many-body systems~\cite{PhysRevLett.132.096501}. 

Notably, in recent years, increasing attention has been focused on the competition between NHSE and localization~\cite{25,26,27,28,29,30,31,32}. The non-reciprocal AAH model can be transformed into Hermitian AAH model by  similar transformations~\cite{26,29,30}, thus exactly obtaining the LEs and localized phase transition points. In the delocalized region, the eigenstates under OBCs are skin states, which can be characterized by the spectral winding number under PBCs. In the Anderson localized region, the localized states are asymmetric with two different LEs. However, unlike the disordered lattice, the wannier-stark lattice has a richer localization behavior~\cite{72,73,74,75,76,77,78,79}. In addition to the ergodic extended states, it also includes weakly ergodic and non-ergodic states~\cite{wei,25}. So far, the fate of the finite Wannier-Stark lattice under non-reciprocity has not been revealed.

In this paper, we analytically and numerically investigate the impact of non-reciprocal non-Hermiticity on the pseudo-mobility edge (ME) and multiple state structures in disorder-free systems. Specifically, we explore how the competition between non-reciprocity and a finite-height Wannier-Stark potential in a 1D   chain affects the system's localization properties. Using transfer matrix and similarity transformation methods, we derive and confirm an analytical expression for the evolution of the pseudo-ME with respect to non-reciprocity. We also identify and analyze the emergence of multiple states under OBCs, including the unique Wannier-Stark-Skin localized state, which exhibits a distinct geometry with hyper-exponential localization on one side and exponential localization on the other. Finally, we present the dynamical fate of these multiple states. Our findings open the door to further exploration of the interplay between non-Hermiticity and single-particle localization.

The structure of this work is as follows. In Sec.~\ref{model}, we introduce the geometry of our model, a finite-height Wannier-Stark potential within non-reciprocal chains. In Sec.~\ref{NHME}, we analytically and numerically present the results on pseudo-MEs. In Sec.~\ref{Multi}, we discuss and analyze the emergence of multiple states in the system. In Sec.~\ref{dynamics}, we provide the dynamical signatures of these multiple states. Finally, we wrap up by comprehensive summarizing and analyzing our findings in Sec.~\ref{conclusion}. 

\section{Model}\label{model}
We consider a 1D   non-reciprocal system with a finite-height Wannier-Stark linear field. The Hamiltonian for this system read as:
\begin{align}
    H =\sum_{j}[t_{l}c^{\dagger}_j c_{j+1} + t_{r}c^{\dagger}_{j+1}c_j]  + \sum_{j}\Delta_j c^{\dagger}_j c_j,\label{eq1}
\end{align}
where, $t_r \equiv e^{-g}$, $t_l\equiv e^{g}$, and
\begin{align}
    \Delta_j = F \cdot j /N.
\end{align}
Here, $c^{\dagger}_{j}$and $c_{j}$ represent the creation and annihilation operators of fermions at site $j$, where $t_r$ and $t_l$ respectively represent the nearest-neighbor (NN) hopping strengths to the right and to the left. $\Delta_j$ represents the on-site potential at site $j$, where $F$ denotes the magnitude of the finite-height Stark potential~\cite{sm}, and $N$ represents the number of lattice sites.  

In this paper, we consider the finite-height potential $F_{\max}$ is finite and independent of the size of the system. Building upon this foundation, we have established an non-Hermitian scenarios which non-reciprocity is induced by modifying the NN hopping term $g \neq 0$ while keeping onsite potential constant.


\section{Non-Hermitian pseudo mobility-edge}\label{NHME}

Under the condition of $g\neq0$, for this non-reciprocal finite height Wannier-Stark model, its Hamiltonian is given by:
\begin{equation}
t_{l}=e^{g}, t_{r}=e^{-g},  \Delta_j= F \cdot j /N,\label{eq3}
\end{equation}
this Hamiltonian can be transformed into the Hermitian finite height Wannier-Stark model through a similarity transformation, and is given by:
\begin{equation}
H^{\prime}=S H S^{-1}=\left(\begin{array}{cccc}
\Delta_1 & t_r & & \\
t_l & \Delta_2 & t_r & \\
& \ddots & \ddots & t_r \\
& & t_l & \Delta_N
\end{array}\right),\label{eq4}
\end{equation}
where $S=\operatorname{diag}\left(e^{-g}, e^{-2 g}, \ldots, e^{-N g}\right)$ is the similarity matrix. For the non-Hermitian matrix $H$, its localization depends on its LE, $ \gamma = \gamma^{\prime} \pm g$ , where $\gamma^{\prime}$ is the LE of the Hermitian matrix $H^{\prime}$ and can be calculated using the transfer matrix method. For the Hermitian matrix $H^{\prime}$, it represents the Hamiltonian of a Hermitian 1D chain system, we let  $\psi_j$ is the wave function at site $j$.  Using the transfer matrix method, we can obtain:
\begin{align}
\binom{\psi_{j-1}}{\psi_j} &=\tilde{T}_{j} \binom{\psi_j}{\psi_{j+1}},\label{eq5}
\end{align}
where $\tilde{T}_{j}$ is:
\begin{align}\tilde{T}_{j} &= \begin{pmatrix}
{E- F j /N} & -1\\
1 & 0
\end{pmatrix}.\label{eq6}
\end{align}
At site $j$, one needs to know the (past) ${\psi_{j-1}}$ value, the (present) values of ${\psi_j}$ and can then compute the (future) value of ${\psi_{j+1}}$ . For simplicity, we denote Eq.~(\ref{eq5}) as:  
\begin{align}
\Psi_j = \tilde{T}_j \Psi_{j-1}.\label{eq7}
\end{align}
Thus, we can express the transfer matrix for transmission across the entire lattice as:
\begin{align}
\binom{\psi_{N+1}}{\psi_N}=\tilde{T}_{N} \tilde{T}_{N-1} \cdots \tilde{T}_{1}\binom{\psi_1}{\psi_0}=\mathbf{Q}_{\mathbf{N}}\binom{\psi_1}{\psi_0},\label{eq8}
\end{align}
where $\mathbf{Q}_{\mathbf{N}}=\prod_{j=1}^N \tilde{T}_{j}$ is a product of matrices, for which the theorem of Oseledec~\cite{46} applies. It states the existence of a limiting matrix:
\begin{align}
\Gamma=\lim _{N \rightarrow \infty}\left(\mathbf{Q}_{\mathbf{N}}^{\dagger} \mathbf{Q}_{\mathbf{N}}\right)^{{1}/{2 N}},\label{eq9}
\end{align}
introducing the eigenvalues ${e^{\gamma_j}}$ and the normalized eigenvectors ${v_j}$ of the symmetric matrix $\Gamma$ one gets:
\begin{align}
\lim _{N \rightarrow \infty}\left(v_j^{\dagger} \mathbf{Q}_{\mathbf{N}}^{\dagger} \mathbf{Q}_{\mathbf{N}} v_j\right)^{{1}/{2 N}}=e^{\gamma_j},\label{eq10}
\end{align}
the LE represents the exponential growth rate of the product of transfer matrices, and is a key observable that reflects the localization properties, from Eq.~(\ref{eq10}) we can get 
\begin{align}
\gamma(j)&= \lim_{N \to \infty} \frac{1}{N} \ln\bigl(\bigl\|\prod_{j=0}^{N-1} \mathbf{Q}_{\mathbf{N}} v_j\bigr\|\bigr),\label{eq11}
\end{align}
where $\bigl\|\cdot \bigl\|$ is the norm of the matrix, the eigenvectors $v_j$ of $\Gamma$ are also eigenvectors of $\mathbf{Q}$, and until now unknown. One might iterate Eq.~(\ref{eq11}) with an arbitrary starting vector $v^0$ instead of $v_j$. The component of $v^0$ leading to the largest LE gets the strongest amplification and Eq.~(\ref{eq11}) converges towards $\gamma_{max}$. In order to obtain all $\gamma$ therefore starts with a unit matrix of initial condition vectors $\psi$. $\psi_0$ is the zero matrix, thus $\left\{u_j^0\right\}=\binom{\mathbf{1}}{\mathbf{0}}$. During the iteration of Eq.~(\ref{eq11}) the vectors $\left\{\tilde{v}_j^0\right\}=\left\{\mathbf{Q}_{\mathbf{N}} v_j^0\right\}$ will lose their orthogonality, using the Gram-Schmidt method the vectors are reorthonormalized after $n$ multiplications:
\begin{align}
v_j=\left(\tilde{v}_j-\sum_{n=1}^{N-1}\left(\tilde{v}_n, \tilde{v}_j\right) \tilde{v}_n\right) /\left\|\tilde{v}_j\right\|,\label{eq12}
\end{align}
while repeating this procedure, applying $n$ transfer-matrix multiplications and then reorthonormalizing, the first vector $v_1$ will converge to the eigenvector corresponding to $\gamma_{max}$, the next vector $v_2$ to the eigenvector of the second largest $\gamma$ and, at the end, the last vector will approach the eigenvector of $\gamma_{min}$. In this way, all eigenvectors ${v_j}$ and LE will be obtained. The introduction of the reorthonormalization steps also solves the problem that a numerical overflow would occurduring the iteration because of the exponential increase of $|| \mathbf{Q}_{\mathbf{N}} v_j^0 ||$. But now Eq.~(\ref{eq11}) is no longer directly applicable. Under the assumption that the vectors $v_j$ have already converged, one can use the norm $b_j= ||\tilde{v_j}||$ and gets:
\begin{align}
\gamma_j = \lim _{N \rightarrow \infty} \frac{1}{N} \sum_{j=0}^{N-1} \ln \left(b_j\right).\label{eq13}
\end{align}
Since $v_j$ is the the eigenvectors of $\mathbf{Q}$, by solving the characteristic equation of the matrix $\tilde{T}_{j}$, as the eigenenergies are in complex form, we can get the LE for non-Hermitian $H$:
\begin{align}
\gamma & \approx \lim _{N \rightarrow \infty} \frac{1}{N} \sum_{j=0}^{N-1} \ln \left(\max \left\{\left|\varepsilon_1\right|,\left|\varepsilon_2\right|\right\}\right)\pm g \nonumber\\
& =\lim _{N \rightarrow \infty} \frac{1}{N} \sum_{j=0}^{N-1}\left|\ln \left(\left|\frac{{-\mu_{r, j}}+\sqrt{\left({-\mu_{r, j}}\right)^2-4 }}{2}\right|\right)\right|\pm g,\label{eq14}
\end{align}
where $\mu_{r, j}={|E|-F j/N}$, $|E|$ represents the norm of the complex eigenenergy, $\varepsilon_1$ and $\varepsilon_2$ are eigenvalues of $\tilde{T}_{j}$, $\varepsilon_1 = 1 / \varepsilon_2$ due to the determinant $|\tilde{T}_{j}| = 1$. Evidently, $\gamma(|E|)=0$ corresponds to the critical energies, since the lattice potential increases linearly, ensuring that the minimum and maximum potentials satisfy $\gamma=0$ is sufficient to guarantee that $\gamma=0$ holds on all lattice sites. In this way, from Eq.~(\ref{eq14}), we can obtain two LEs, and the localization phase transition is determined by the smaller LE. Therefore, can obtain the critical energy as:
\begin{align}
    |E| = \left\{ 
    \begin{array}{lc}
        t_r+t_l, \\
        (t_l-t_r)\sqrt{{1-(F_{{max}})^2/(t_r+t_l)^2}},\\
        F_{{max}}-t_r-t_l,\\
        t_r+t_l-F_{{max}}.
    \end{array}
    \right.\label{eq15}
\end{align}
\begin{figure}[bt]
\hspace*{-0.5\textwidth}
\includegraphics[width=0.5\textwidth]{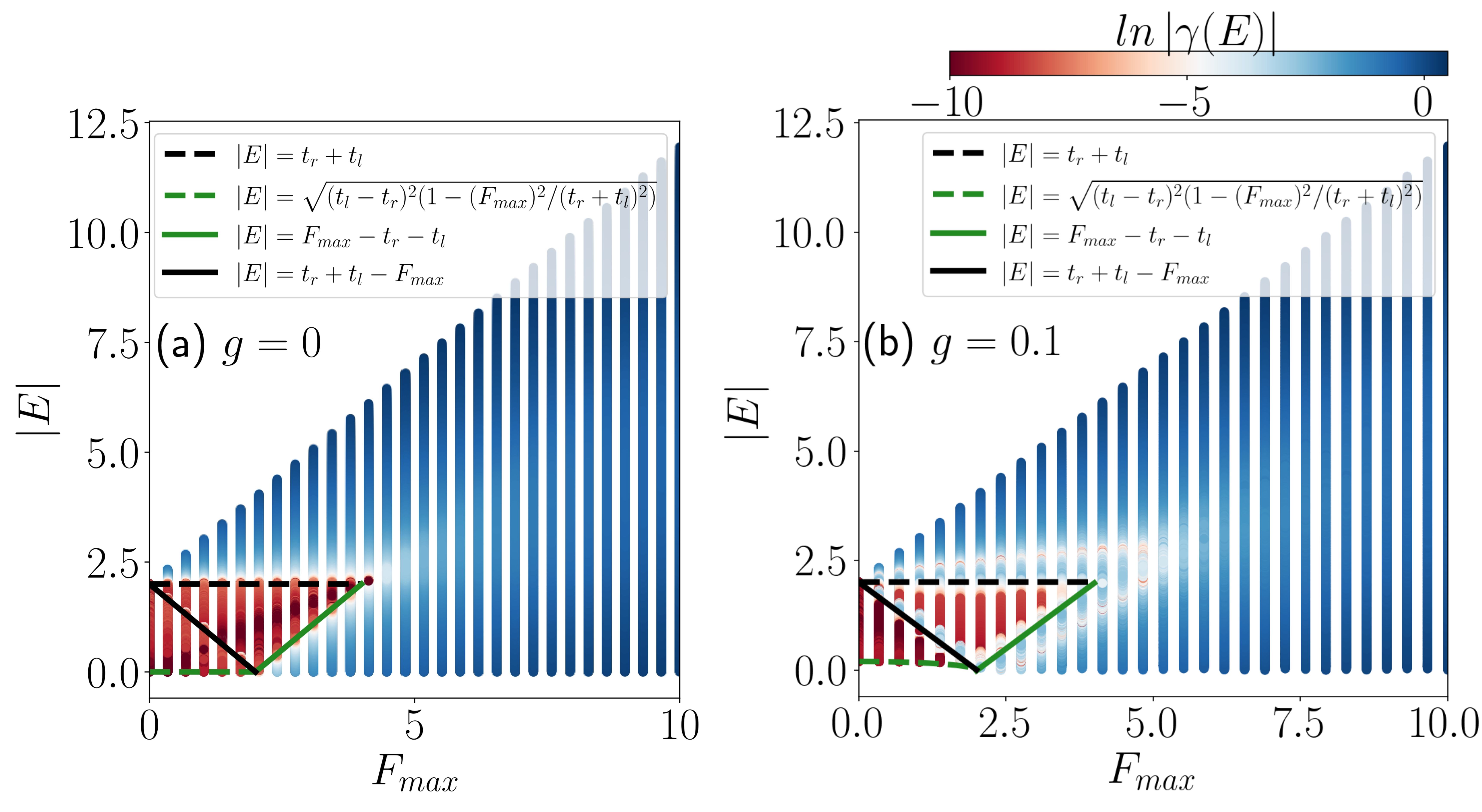} 
\caption{Spectra and critical energies as a function of $F_{max}$. In the figures, $g$ represents the non-reciprocal strength, and the color represents the value of $\ln |\gamma(E)|$. We employ a finite-height Wannier-Stark potential, which is far from meeting the requirements for strong Wannier-Stark localization. The red (blue) color regions correspond to the ergodic (weakly ergodic) regions, respectively. The green and black dashed and solid lines mark the critical energies separating ergodic states from weakly ergodic states. Here, $N=5000$. \label{fig1}}
\end{figure}
\begin{figure*}[bt]
\hspace*{-1\textwidth}
\includegraphics[width=1\textwidth]{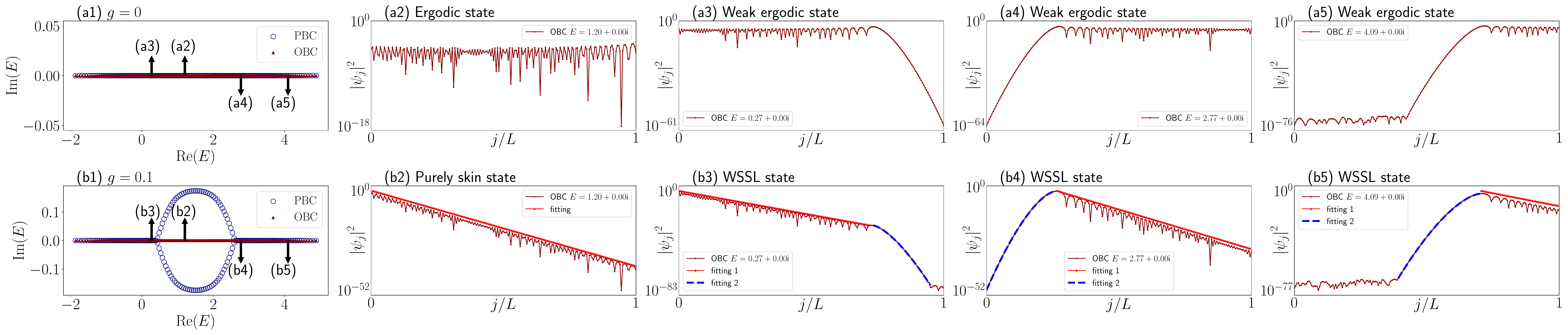} 
\caption{Wave function plots for different energy levels under OBC. Imaginary energy versus real energy plots in Hermitian system (a1), non-Hermitian system with non-reciprocity (b1)   under PBC and OBC. To highlight the impact of non-reciprocity on the wave function states, Group (a) shows the Hermitian system, while Group (b) shows the non-Hermitian system with added non-reciprocity. In addition, in non-Hermitian systems, we selected one energy level on the left side of the energy loop , one energy level inside the energy loop, and two energy levels on the right side of the energy loop to obtain different states at various energy levels within the system. The red and blue lines represent the fits for the skin state and the hyperexponential localized state, respectively. The fitting equations are given by  Eq.~(\ref{eq23}) and  Eq.~(\ref{eq25}). Here, $N=200, F_{max}=3 $.}\label{fig3}
\end{figure*}

To verify the analytical results, we employed the following method to calculate the LE and numerically determined its critical energy. Starting from the original definition of the LE, we can obtain:
\begin{align}
\gamma & =\lim _{N \rightarrow \infty} \frac{1}{N} \ln \left(\left|\Psi_{N-1}\right| /\left|\Psi_0\right|\right) \nonumber\\
& =\lim _{N \rightarrow \infty} \frac{1}{N} \ln \bigl(\frac{\left|\Psi_{N-1}\right|}{\left|\Psi_{N-2}\right|} \frac{\left|\Psi_{N-2}\right|}{\left|\Psi_{N-3}\right|} \ldots \frac{\left|\Psi_1\right|}{\left|\Psi_0\right|}\bigl) \nonumber\\
& =\lim _{N \rightarrow \infty} \frac{1}{N} \sum_{j=0}^{N-2} \ln \bigl(\frac{\left|\Psi_{j+1}\right|}{\left|\Psi_j\right|}\bigl), \label{eq16}
\end{align}
here $\left|\Psi_j\right|=\sqrt{\left|\psi_{j+1}\right|^2+\left|\psi_j\right|^2}$ is the norm of the vector. In both the non-reciprocal non-Hermitian and onsite dissipation potential cases, the LE is computed over all lattice points. The magnitude of the LE can effectively separate ergodic states from other states. The specific procedure involves selecting a normalized initial state first:
\begin{align}
\Psi_0=\binom{1}{0},\label{eq17}
\end{align}
and setting $ \gamma(E) = 0 $ as the initial LE value. Next, by multiplying the wave function with the transfer matrix Eq.~(\ref{eq5}), we obtain a new wave function:
\begin{align}
    \Psi_{j+1}=T_{j+1} \Psi_j,\label{eq18}
\end{align}
subsequently, normalize this new wavefunction and then compute the LE from it:
\begin{align}
    \gamma=\gamma+ \frac{1}{N} \ln \left(\left|\Psi_{j+1}\right| /\left|\Psi_j\right|\right),\label{eq19}
\end{align}
finally, iterate the above steps from lattice point $0$ to $ N-2 $ to calculate the final LE. Here, we plot both the analytical and computed results in FIG.~\ref{fig1} We characterize the critical energy using the LE obtained through Eq.~(\ref{eq19}), during the computation, the LE is obtained with respect to the complex energy $ E_C $ using the $\bigl\|\cdot \bigl\|$ method to handle the wave function and transmission matrix. Therefore, in FIG.~\ref{eq1}, we take the modulus of the complex energy to be $ |E| $. In FIG.~\ref{eq1} (a)-(b), we consider the case of a finite-height Wannier-Stark linear field with non-reciprocal. Their critical energies are denoted as $|E|=t_r+t_l$, $|E|=(t_l-t_r)\sqrt{{1-(F_{{max}})^2/(t_r+t_l)^2}}$, $|E|=F_{{max}}-t_r-t_l$,  $t_r+t_l-F_{{max}}$  which align well with Eq.~(\ref{eq15}). It is worth mentioning that the critical energy we derived is in the thermodynamic limit, due to size effects, a small portion of the ergodic states traversing the mobility edge will appear outside the phase boundary in FIG.~\ref{fig1}. For details, see Appendix ~\ref{B}.

In FIG.~\ref{fig1}, it is evident that, as the non-reciprocal strength $g$ increases, the ergodic state region gradually contracts. This trend is also observed with the potential $F_{max}$, where an increase in $F_{max}$ leads to a similar effect. Specifically, from FIG.~\ref{fig1} (a)-(b), it can be observed that when $F_{max}$ increases to $4$, there are no longer any ergodic states in the system, all states transition to weakly ergodic states, indicating that the non-reciprocal strength $g$ does not influence the potential's control over the ergodic states.

\section{The multiple states\label{Multi}}

In classical Anderson disordered systems, there exist extended (ergodic) states and Anderson localized states. Under the influence of non-reciprocity, the extended states become skin states localized at the boundaries under open boundary conditions, while the Anderson localized states remain localized at their original positions but exhibit different LEs on the left and right sides. However, for systems with Wannier-stark localization, there are more diverse states:  ergodic, weakly ergodic, and non-ergodic states, and building on this foundation, we introduce non-reciprocity to explore the changes in the states within the system. The Hamiltonian of non-reciprocal finite height Wannier-Stark model can be derived from a similarity transformation and is given by Eq.~(\ref{eq4}). For the Hermitian Hamiltonian $H^{\prime}$, the localization transition point is $t=\sqrt{t_lt_r}=1$. Let $\psi^{\prime}$ be an eigenstate of the Hamiltonian $H^{\prime}$; then, the eigenstate $\psi$ of the Hamiltonian $H$ satisfies $\psi=S^{-1} \psi^{\prime}$. Thus, for an extended eigenstate of the Hamiltonian $H^{\prime}, S^{-1}$ causes the wave function to be exponentially localized on the left (right) boundary for $g>0(g<0)$, which results in non-Hermitian skin effects. For a skin state, the corresponding wave function is given by:
\begin{align}
\left|\psi_j\right| \propto e^{-\gamma{j}} ,\label{eq20}
\end{align}
where $\gamma$ is the LE for Hamiltonian $H$ and can be expressed as:
\begin{align}
\gamma=\ln \left(\frac{t_l}{t_r}\right) .\label{eq21}
\end{align}
 The hyper-exponential localized states formed by Wannier-Stark localization are given by:
\begin{align}
\left|\psi_j\right| \propto \begin{cases}e^{|\psi_{j_0}|^2-\alpha\left(j-j_0\right)^\beta}, & j>j_0, \\ e^{|\psi_{j_0}|^2-\alpha\left(j_0-j\right)^\beta}, & j<j_0, \end{cases}\label{eq22}
\end{align}
where $|\psi_{j_0}|^2$ is the density of wave function at localization center.
In FIG.~\ref{fig3}, we have plotted the wave functions for the Hermitian system  (a2-a5) and the wave functions for the non-Hermitian system caused by the non-reciprocal term (b2-b5) at different energy levels. In FIG.~\ref{fig3} (b2), the wave function localized in a purely skin state given by Eq.~(\ref{eq20}). In FIG.~\ref{fig3} (b3), the skin state is still described by Eq.~(\ref{eq20}), however, due to the presence of the skin state, the localization center of the hyper-exponential localized (HEL) state shifts to the rightmost end of the skin state $j_1$, and Eq.~(\ref{eq22}) transforms into:
\begin{align}
\left|\psi_j\right| \propto e^{|\psi_{j_0}|^2-\alpha\left(j-j_1\right)^\beta}.\label{eq23}
\end{align}
In FIG.~\ref{fig3} (b3), the localization center $j_1=0.7337$ and  $\alpha=472.5835 , \beta=1.46636$. In FIG.~\ref{fig3} (b4), the localization center $j_0$ of the skin state is not at the boundary, since the Wannier-Stark localized state is located to the left of the skin state and shares the same localization center with it, the description of the wave functions for both the skin state and the hyper-exponential localized state transforms into:
\begin{align}
\left|\psi_j\right| \propto\begin{cases} e^{-\gamma{({j}-j_0)}} ,& {j>j_0},\\
 e^{|\psi_{j_0}|^2-\alpha\left(j-j_0\right)^\beta},&{j<j_0},  \end{cases}\label{eq24}
\end{align}
and the fitting data for HEL state is $j_0=0.2613,\alpha=427.4325, \beta =1.60094$. In FIG.~\ref{fig3} (b5), since $g > 0$, our skin state is directed to the left, causing the HEL state on the left to share a localization center $j_0$ with the skin state. The localization pattern of the wave functions in (b5) can be derived from Eq.~(\ref{eq20}) and~(\ref{eq22}) 
like this: 
\begin{align}
\left|\psi_j\right| \propto\begin{cases} e^{|\psi_{j_0}|^2-\alpha\left(j_0-j\right)^\beta},& {j<j_0},\\
e^{-\gamma{({j}-j_0)}} ,& {j>j_0},\\
\end{cases}\label{eq25}
\end{align}
the fitting data for the HEL state on the left side is: $j_0=0.7035, \alpha= 417.4638, \beta=1.58369$. 
\begin{figure}[H]
\hspace*{0\textwidth}
\includegraphics[width=0.5\textwidth]{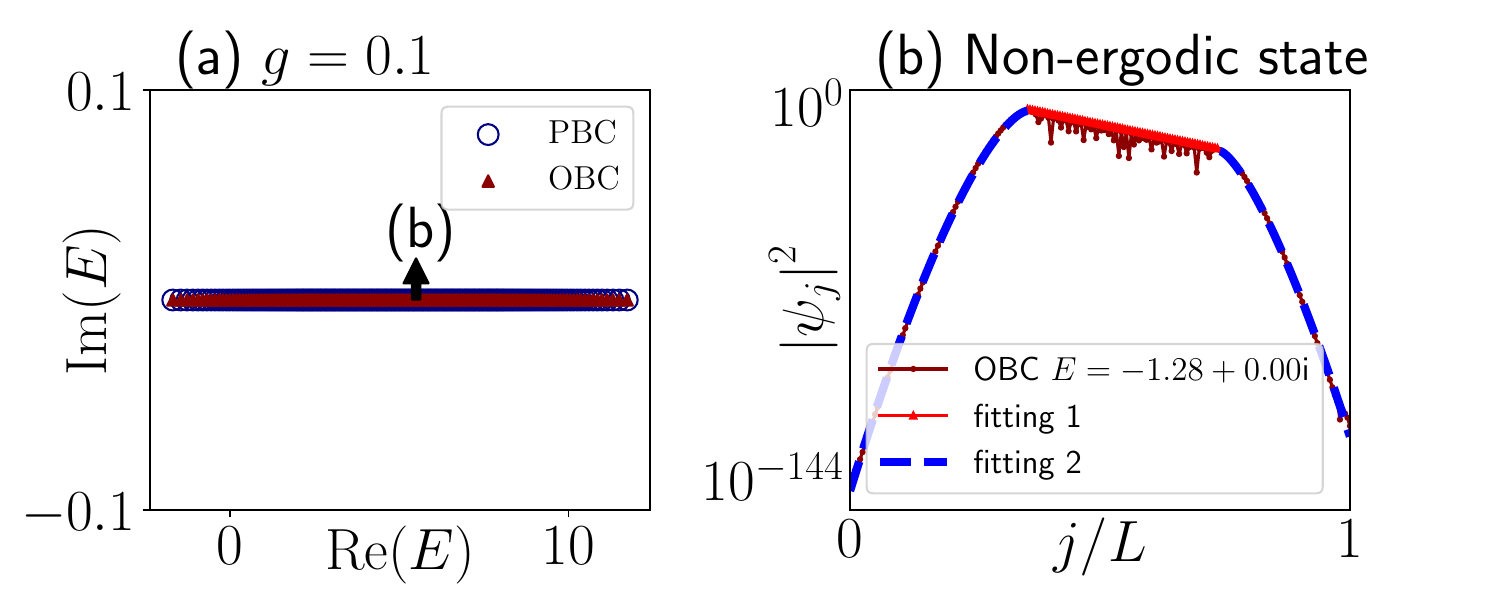} 
\caption{Imaginary energy versus real energy plots in non-Hermitian system with non-reciprocity under PBC and OBC. Wave function plot for non-ergodic state under OBC. The red and blue lines represent the fits for the skin state and the hyper-exponential localized state, respectively. Here, $F_{max} = 10$ and $N = 200$. }\label{fig2}
\end{figure}
\begin{figure}[bt]
\hspace*{-0.5\textwidth}
\includegraphics[width=0.5\textwidth]{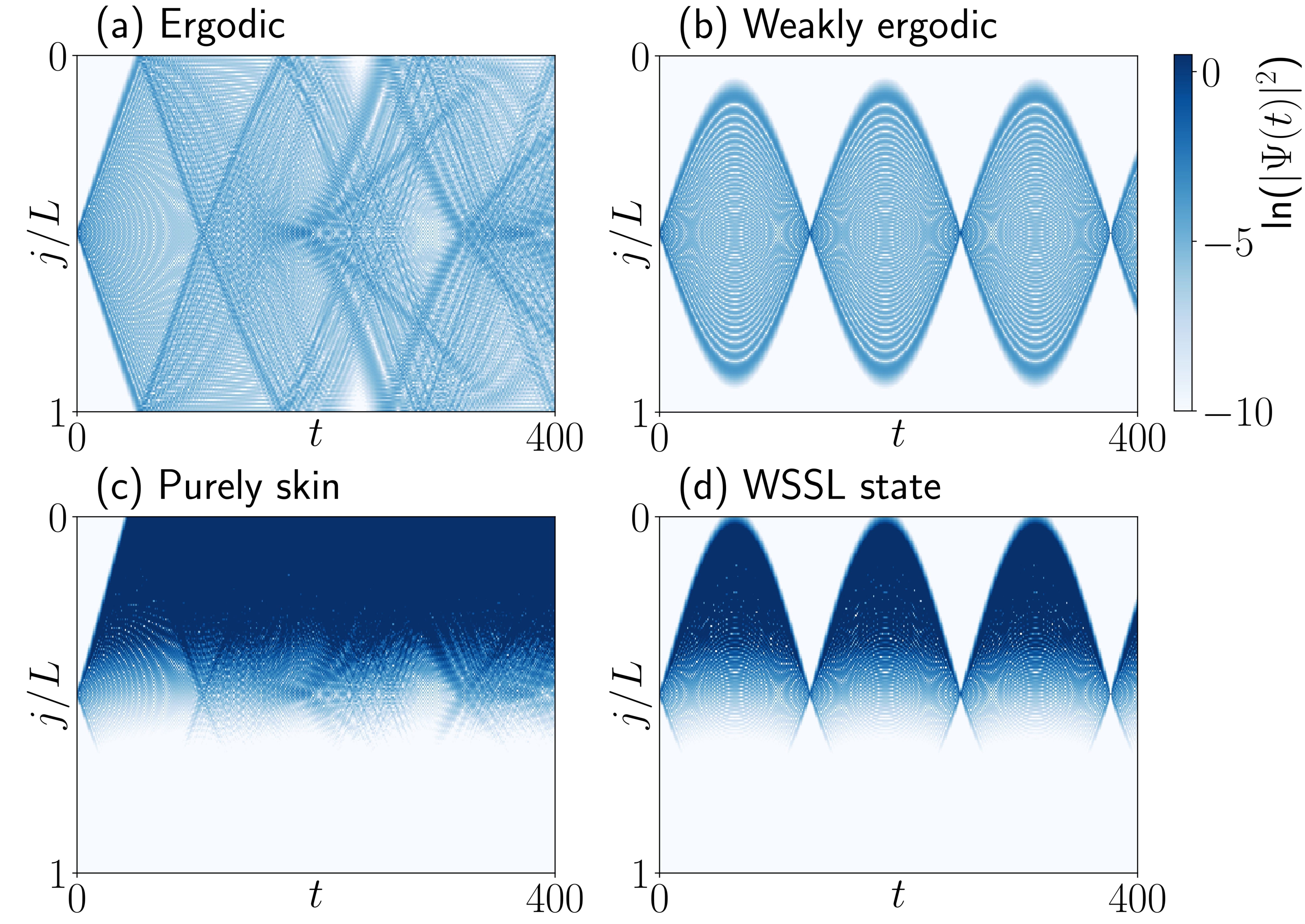} 
\caption{Time evolution of the wave function $|\Psi(t)|^2$ for (a) and (b) in Hermitian system, (c) and (d) in non-Hermitian system with non-reciprocal strength $g = 0.1$ under OBC. In panels a and c, we set the Wannier-Stark ladder potential height to 3, while in panels b and d, to make the Wannier-Stark localization more visually prominent, we increased the Wannier-Stark ladder potential height to 10. The darker (lighter) colors represent higher (lower) state density. Here, $N = 200$. }\label{fig4}
\end{figure}
In FIG.~\ref{fig2}, we demonstrate a more localized state by increasing $F_{max}=10$. As shown, with the increase of $F_{max}$, the energy loop under PBC disappears, and the system no longer exhibits ergodic states. This enhanced localized state consists of two types of localization: hyper-exponential localization and skin effect localization. Due to the directionality of the skin effect, the center of the hyper-exponential localized state on the left side of the skin state is located at $j_0=0.355$, sharing a common localization center with the skin state. On the right side of the skin state, the center of the hyper-exponential localized state is at $j_1=0.74$, marking the furthest extent of the skin state. The localization modes of these states can be represented as: 
\begin{align}
\left|\psi_j\right| \propto\begin{cases} e^{|\psi_{j_0}|^2-\alpha\left(j_0-j\right)^\beta},& {j<j_0},\\
e^{-\gamma{({j}-j_0)}} ,& {j_0<j<j_1},\\
e^{|\psi_{j_1}|^2-\alpha\left(j-j_1\right)^\beta},& {j>j_1},\\
\end{cases}\label{eq30}
\end{align}
the fitted data for the HEL states on the left and right sides are respectively: $\alpha= 707.4277, \beta=1.51528$, and $\alpha= 734.28618, \beta=1.40524$. 

It can be observed that under OBC, influenced by non-reciprocity, the states at the same energy transition from ergodic or weakly ergodic states (shown as complete or partial plateaus in (a2-a5) to skin states (shown as sloped plateaus in (b2-b5)). By comparing FIG.~\ref{fig3} (a2) and (b2), we can see that under the influence of non-reciprocity, the original ergodic state forms topological point gaps in the complex plane, and the corresponding eigenstates are localized at the boundaries, forming a purely skin state. Comparing FIG.~\ref{fig3} (a3-a5) and (b3-b5), these weakly ergodic states transition into a novel WSSL state. It is not difficult to understand that at a certain energy, the Wannier-Stark localization caused by the ladder potential competes with the skin effect brought by the non-reciprocal term, forming this novel WSSL state. Moreover, due to the different positions of the Wannier-Stark localization, the localization centers of these novel WSSL states are also different. On the one hand, on the  side of the localization center, Wannier-Stark localization causes hyper-exponential localization. On the other hand, the addition of the non-reciprocal term transforms the original ergodic state into an exponentially localized state caused by the skin effect, appearing on the other side of the localization center.

\section{Fingerprints of dynamics\label{dynamics}}
In this section, we dynamically investigate the dynamical fingerprints of the multiple states in the Hamiltonian Eq.~(\ref{eq1}). We initially encode one particle located in the center site $n_0$ of the lattice. The time evolution states are determined by 
\begin{align}
    \ket{\Psi_t}=\frac{e^{-\text{i}tH}\ket{\Psi_0}}{||e^{-\text{i}tH}\ket{\Psi_0}||}.\label{eq26}
\end{align}
The particle density for any site and time $t$ is
\begin{align}
    |\Psi^n_t|^2=|\langle n\ket{\Psi_t}|^2.\label{eq27}
\end{align}Where, $\ket{n}$ is $n-th$ computational basis of the Hilbert space. In FIG.~\ref{fig4}, we present the distribution of state densities under OBC as they evolve over time. First, we set the $g=0$, for a  weak   potential $F_{max}=3$ in FIG.~\ref{fig4} (a), when the initial state energy is in the ergodic region, the wave function spreads across the entire chain during time evolution. In FIG.~\ref{fig4} (b), as the potential increases to $F_{max}=10$, the wave function consists predominantly of Wannier-Stark localized states under Bloch oscillations, with a small portion of ergodic states, which is referred to as a weak ergodic state. Then we let $F_{max}=3$ and $g=0.1$ in FIG.~\ref{fig4} (c), it is evident that the initial ergodic state rapidly transitions to purely skin state. Next, we set $g=0.1$, for $F_{max}=10$, it can be clearly observed that, as time evolves, the wave function gradually accumulates towards the boundary and exhibits a pronounced skin effect, and the weak ergodic state formed by the Wannier-Stark potential does not disappear. By connecting this with FIG.~\ref{fig3}, we can derive this novel WSSL state is composed of states in FIG.~\ref{fig3}(b3-b5) and tends to evolve more towards the side with greater non-reciprocity. Additionally, the state FIG.~\ref{fig3}(b3) contributes in a form similar to a skin mode, all these reasons resulting in a novel WSSL state in FIG.~\ref{fig4} (d).  From the results in FIG.~\ref{fig4}, it can be seen that The inclusion of non-reciprocal terms only affects the ergodic state, transforming it from the initial ergodic state to a skin localized state. These results further demonstrate that this novel WSSL state emerges under appropriate ladder potential height and non-reciprocal strengths. There are two types of localization in it: Wannier-Stark localization and localized states formed by the skin effect.

\section{Conclusion}\label{conclusion}
In summary, we conducted both analytical and numerical studies on the pseudo MEs and wave functions in a 1D chain with a finite-height Wannier-Stark linear field, under the influence of non-reciprocal non-Hermitian modulation. We employed the analytic transfer matrix method to calculate the system's LE and identified the critical energies that differentiate weakly ergodic states from fully ergodic states based on the LE. 

On the one hand, we provide the exact pseudo-ME under non-reciprocal modulation. The results show that the exact pseudo-ME in the system is modulated by the strength of non-reciprocity, regardless of the presence of a   potential. On the other hand, we thoroughly discuss the regulation of multiple states in the system through the lens of similarity transformations, with a segmented analysis of the impact of both the finite-height Wannier-Stark potential and non-reciprocity on wave function restructuring. We find that due to the presence of the skin effect, the originally ergodic regions of the wave function are compressed into an exponential decay. Moreover, distinct states coexisting with both exponential and hyper-exponential localization emerge, which is in stark contrast to non-reciprocal disordered non-Hermitian systems. Additionally, we present the wave packet dynamics, revealing the dynamical signatures of the emergent multiple states.

Our findings not only deepen the understanding of the interplay between non-Hermicity and disorder-free localization but also suggest that such phenomena could be realized and studied experimentally in platforms, $e. g.$, photonic lattices~\cite{52,53}, cold atomic gases in optical potentials~\cite{54,55}, or electrical circuits~\cite{56,57,58}, where the effects of non-Hermiticity and finite-height Wannier-Stark linear field can be engineered and precisely controlled. 

\vspace{8pt}

\begin{acknowledgments}
JXZ acknowledges the Shanghai University Distinguished Professor Research Start-up Project, the National Natural Science Foundation of China (Grant No.12374046 and No.11874316), the National Basic Research Program of China (Grant No.2015CB921103), and the Program for Changjiang Scholars and Innovative Research Team in University (Grant No.IRT13093).
\end{acknowledgments}

\appendix
\section{Differences between the finite and infinite high Stark potential used in this work\label{A}}
\begin{figure*}[bt]
\hspace*{-0.9\textwidth}
\includegraphics[width=0.9\textwidth]{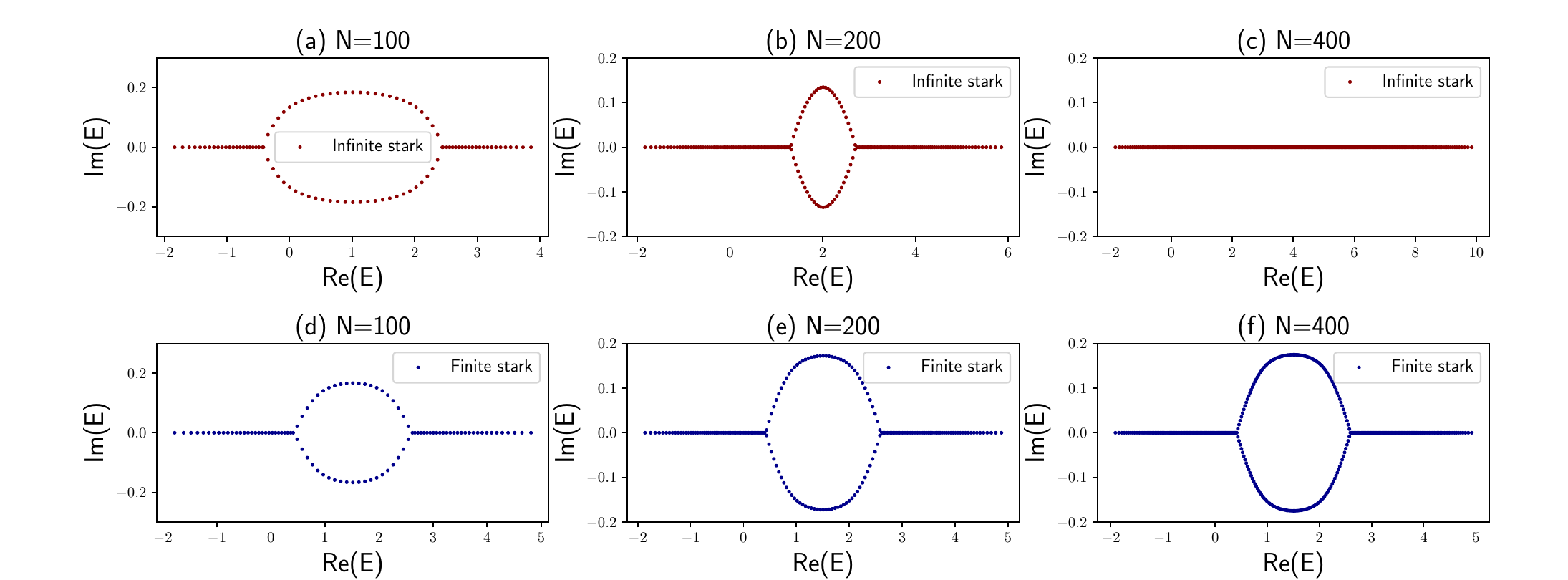} 
\caption{ Diagrams of real and imaginary energies under PBC for infinite and finite high Stark potential systems with non-reciprocal $g=0.1$. The red scatter plots (a), (b), (c) represent the infinite height system with a step height of 0.02, and the blue scatter plots (d, e, f) represent the finite height system with a maximum potential of 3. The system sizes are set to $N=100,200,400$ from left to right. We can clearly see that in the infinite height Stark potential system, as the system size 
$N$
 increases, the energy loop gradually disappear. Conversely, in the finite height Stark potential system, the energy loop persist, does not change with size.
}\label{figA2}
\end{figure*}

Here, we present the differences in the energy spectra of finite and infinite height Stark potentials. As shown in the FIG~\ref{figA2}, the energy spectrum of the infinite height Stark potential system is affected by the system size, whereas the spectrum of the finite height Stark potential system is not size-dependent. In a true infinite height Stark potential system, ergodic states do not exist, while the finite height Stark potential system exhibits a richer variety of states.

\section{The effect of finite size on the ergodic region\label{B}}
\begin{figure*}[bt]
\hspace*{-0.9\textwidth}
\includegraphics[width=0.9\textwidth]{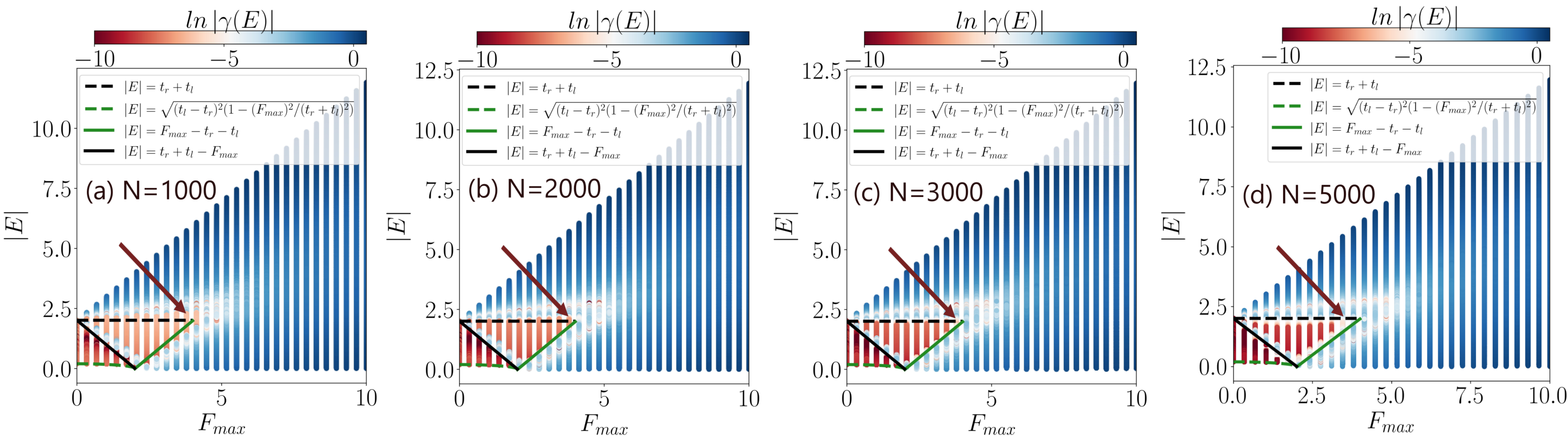} 
\caption{The energy spectrum for $g=0.1$
 at different sizes $N$. The color represents the value of $\ln |\gamma(E)|$. We employ a finite-height Wannier-Stark potential, which is far from meeting the requirements for strong Wannier-Stark localization. The red (blue) color regions correspond to the ergodic (weakly ergodic) regions, respectively. The green and black dashed and solid lines mark the critical energies separating ergodic states from weakly ergodic states. Here, from (a)-(d),$N=1000,2000,3000,5000$, respectively.
}\label{figA1}
\end{figure*}

In Sec.~\ref{NHME}, we point out that a small portion of ergodic states crossing the pseudo mobility edge is caused by finite size effects. In FIG~\ref{figA1}, we show the spectra of different size systems under the same non-reciprocal strength $g=0.1$. The system sizes for (a, b, c, d) are $N
=1000,1500,2000,3000$ respectively. It can be observed that as $N$ increases, the mobility edge of the weak ergodic region gradually aligns with the analytical solution.


\bibliography{bibliography}

\begin{thebibliography}{71}
\bibitem [{sm()}]{sm}%
  \BibitemOpen
  \href@noop {} {}\bibinfo {note} {Please note that in theoretical contexts, the term ``infinite height Wannier-Stark potential" often refers to a system studied in the thermodynamic limit, where the system size tends to infinity, and the potential is assumed to have no explicit boundaries. However, in a finite system, the Wannier-Stark ladder is truncated because only a limited number of states are physically available. Moreover, experimental setups often impose a maximum potential energy or cutoff, effectively making the potential ``finite height."}\BibitemShut {NoStop}%
\end{thebibliography}


\end{document}